\documentclass{svjour2}                    
\smartqed  
\usepackage{graphicx}
\usepackage{graphicx}
\usepackage{epstopdf}
\usepackage{dcolumn}
\usepackage{bm}
\journalname{Acta Mech}
\begin{document}

\title{Effect of boundary vibration on the frictional behavior of a dense sheared granular layer
}

\author{B. Ferdowsi         \and\\
        M. Griffa \and\\
        R.~A. Guyer \and \\
        P.~A. Johnson \and\\
        J. Carmeliet
}

\institute{B. Ferdowsi \at
	Swiss Federal Institute of Technology Z\"{u}rich (ETHZ) - DBAUG, Schafmattstrasse 6, CH-8093 Z\"{u}rich, Switzerland and Swiss Federal Laboratories for Materials Science and Technology (Empa),
	ETH Domain - \"{U}berlandstrasse 129, CH-8600, D\"{u}bendorf (Z\"{u}rich), Switzerland
              \email{behrooz.ferdowsi@empa.ch, behroozf@ethz.ch}           
           \and
           M. Griffa \at
           Swiss Federal Laboratories for Materials Science and Technology (Empa),
	ETH Domain - \"{U}berlandstrasse 129, CH-8600, D\"{u}bendorf (Z\"{u}rich), Switzerland
	\and
	R.~A. Guyer \at
	Solid Earth Geophysics Group, Los Alamos National Laboratory - MS D443, NM 87545, Los Alamos, USA and
	Department of Physics, University of Nevada, Reno (NV), USA
	\and
	P.~A. Johnson \at
	Solid Earth Geophysics Group, Los Alamos National Laboratory - MS D443, NM 87545, Los Alamos, USA
	\and
	J. Carmeliet \at
	Swiss Federal Laboratories for Materials Science and Technology (Empa),
	ETH Domain - \"{U}berlandstrasse 129, CH-8600, D\"{u}bendorf (Z\"{u}rich), Switzerland and Chair of Building Physics, Swiss Federal Institute of Technology Z\"{u}rich (ETHZ) - Wolfgang-Pauli-Strasse 15, CH-8093,
	Z\"{u}rich, Switzerland
}

\date{Received: 29 January 2013 / Accepted: 29 May 2013} 

\maketitle

\begin{abstract}
We report results of 3D Discrete Element Method (DEM) simulations aiming at investigating the role of the boundary vibration in inducing frictional weakening in sheared granular layers. We study the role of different vibration amplitudes applied at various shear stress levels, for a granular layer in the stick-slip regime and in the steady-sliding regime. Results are reported in terms of friction drops and kinetic energy release associated with frictional weakening events. We find that larger vibration amplitude induces larger frictional weakening events. The results show evidence of a threshold below which no induced frictional weakening takes place. Friction drop size is found to be dependent on the shear stress at the time of vibration. A significant increase in the ratio between the number of slipping contacts to the number of sticking contacts in the granular layer is observed for large vibration amplitudes. These vibration-induced contact rearrangements enhance particle mobilization and induces a friction drop and kinetic energy release. This observation provides some insight into the grain-scale mechanisms of frictional weakening by boundary vibration in a dense sheared granular layer. In addition to characterizing the basic physics of vibration induced shear weakening, we are attempting to understand how a fault fails in the earth under seismic wave forcing. This is the well know phenomenon of dynamic earthquake triggering.  We believe that the granular physics are key to this understanding.

\end{abstract}

\section{Introduction}
Granular materials are made up of many distinct grains that often interact with each other through dissipative contact forces. They are abundant in daily life and their bulk behavior spans a broad range of states unlike that of either solids or liquids. Among these behaviors is the transition from a solid-like to a fluid-like behavior or vice versa. This transition is the basis of a broad spectrum of natural and geophysical processes as well as industrial applications. In particular, earthquake initiation on mature faults that contain a granular fault gouge (as a result of wear and frictional slip of the fault interfacial surfaces) is attributed to this solid-to-fluid-like transition.  A fault system accumulates strain energy during inter-seismic periods, known as the ``stick'' phase, and a ``slip event'' corresponds to an earthquake\cite{Brace1966,Brace1970,Johnson1973}. The stick-slip dynamic regime of a sheared granular layer is controlled by mechanical and physical properties of the layer, including its confining pressure and shearing velocity\cite{Daniels2008,Hayman2011}. The intrinsic stick-slip dynamics of a granular layer can be perturbed by external factors including boundary vibrations. Laboratory scale observations as well as Discrete Element Method (DEM) simulations show and confirm that mechanical and acoustic vibrations with adequate amplitudes can change the mechanical and frictional properties of a confined and sheared granular layer, and consequently its macro-scale response~\cite{Luding1994,Savage2008,Janda2009,Capozza2009,Melhus2009,Jia2011,vanderElst2012}. Many aspects of this vibration-induced changes including its grain-scale mechanisms, its dependence on the loading state of the granular layer and on the vibration amplitude are unexplored. Understanding and characterizing the effects of boundary vibration is of importance since at a larger scale, numerous observations show that seismic waves, radiated by an earthquake can trigger earthquakes at other mature faults both near and far away from the original one~\cite{Pollitz2012,Shelly2011,Velasco2009,Marsan2008,Gomberg2004,Gomberg2001}. This observed phenomenology is termed Dynamic Earthquake Triggering (DET)~\cite{Freed2005}, and its physical origin is thought to be related to the frictional evolution of the granular fault gouge layer.

We recently investigated the role of boundary vibration on slip triggering by analyzing the affine and non-affine deformation fields (\cite{Falk1998,DiDonna2005}) inside the granular layer and their spatial-temporal evolution\cite{Griffa2011,Griffa2012}. The vibration amplitude is shown to have significant influence on the size of triggered slip events in a 2D DEM model of a sheared granular layer~\cite{Griffa2013}.

Here, we will present results of a 3D DEM numerical simulation of a confined sheared granular layer subjected to boundary vibration. The goal of this investigation is to understand the influences of boundary vibration amplitude and the time location of its application on the immediate frictional weakening of the granular layer. To this end, we have applied vibration at various temporal locations during the granular layer dynamics. In one case, the dynamics is of stick-slip type. In the second case, the granular layer is steadily sliding. The paper is structured as follows: first, we introduce our numerical modeling approach and the model geometry. Then, an example of vibration-induced frictional weakening is given and the grain contact evolution of the granular layer is studied in correspondence of the friction drop induced by vibration. The influences of the vibration amplitude and the shear stress level at which the vibration is applied on the size of the frictional weakening event are investigated finally.  

\section{Model description}
\label{sec:model}
\begin{figure}
\includegraphics[width=0.60\textwidth]{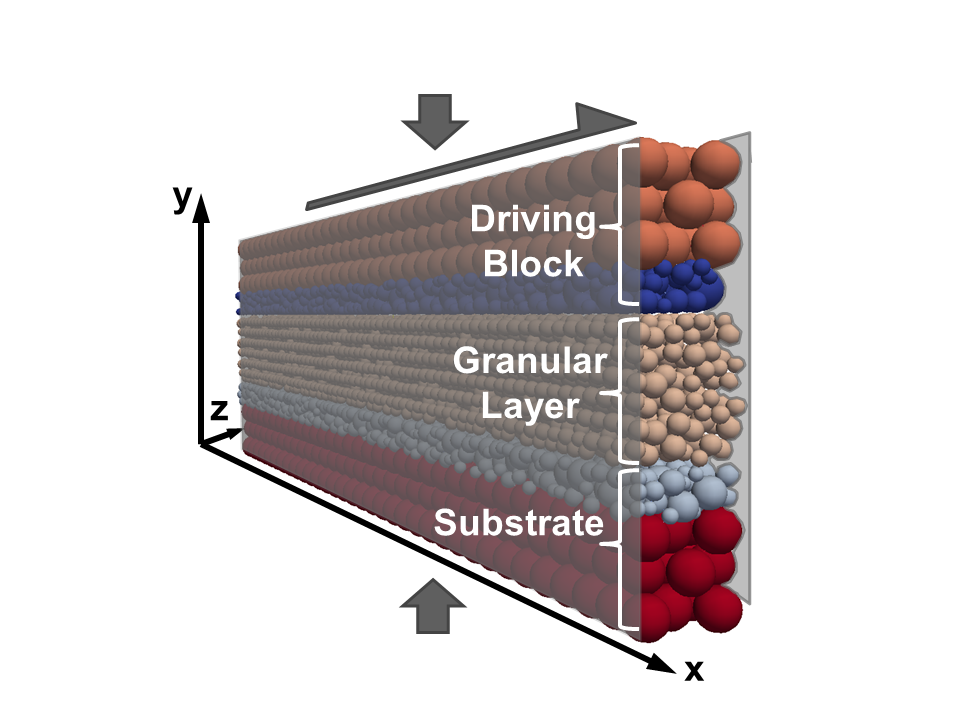}
\caption{3D DEM model made of the driving block (top), granular gouge layer (center) and the substrate block (bottom). The particle colors reflect different physical components of the layer.}
\label{fig:fig1} 
\end{figure}

Figure~\ref{fig:fig1} shows the geometry of the 3D DEM model fault gouge layer. The model consists of three sets of particles: a top driving block, a mirroring
substrate block and a granular gouge layer. The driving and substrate blocks are used to confine the granular gouge by applying a constant normal force in the $Y$-direction. The top set of particles of the driving block move with a
prescribed velocity in the positive $X$-direction and applies a shearing force to the granular gouge layer. 

Each variable/parameter in our 3D DEM model is expressed in terms of the following basic dimensional units: 
$L_0 = 150\ \mu m$, $t_0 = 1\ s$ and $M_0 = 1\ kg$, for length, time and mass, respectively. $L_0$ represents the largest particle radius within the overall DEM model. 

The driving and substrate blocks are modeled as a system of spherical bonded particles, and are each comprised of two distinct layers. The first layer (top layer - brown colored particles- 
for the driving block and bottom layer -red colored particles- for the substrate block) consists of a Hexagonal Close Packed (HCP) arrangement of monosized particles with radius $L_0$, while the second layer (roughness layer, dark and light blue colored particles) consists of 
particles with radii distributed within $[0.3; 1.0]L_0$. The particle assemblies of the roughness layers were initially generated using a space-filling particle insertion method~\cite{Schoepfer2009}. The driving and substrate blocks have thickness of approximately 7.0${L_0}$ ($ \frac{Thickness_{roughness \; layer}}{Thickness_{HCP \; layer}} = 0.32$). The HCP layers are arranged to produce adequate rigidity for the interaction of the driving and substrate blocks with the granular gouge layer. The roughness layers facilitate stick-slip dynamics by increasing the interaction of the granular gouge with the driving and substrate blocks. In addition, the structure of the driving (substrate) block allows for its dynamic interaction with the granular layer during shearing, a similar role to that played by tectonic bounding blocks in a fault system.

\begin{figure}
\includegraphics[width=0.55\textwidth]{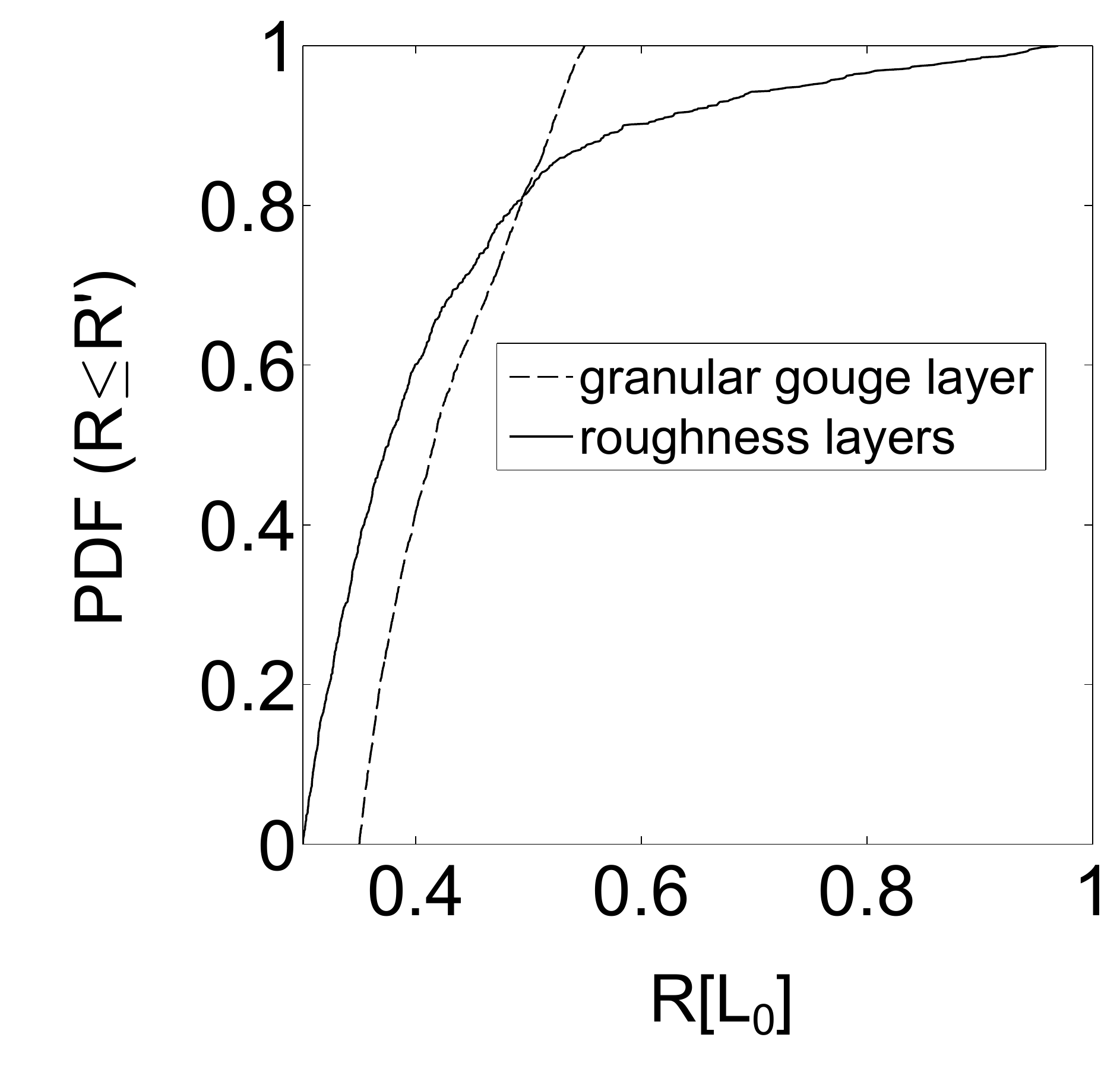}
\caption{Particle Size Distribution of the roughness layer of driving block and substrate and the granular gouge layer.}
\label{fig:fig2}
\end{figure}

The granular gouge layer includes a set of spherical, unbonded particles and its initial assembly is generated using the same space-filling particle insertion method used for the driving and substrate blocks~\cite{Schoepfer2009}. The radius of the granular gouge particles varies from 0.35$L_0$ to 0.55$L_0$ and corresponds to the size range, $[105; 150]\mu{\it m}$, of the silica glass beads used as model fault gouge in laboratory shear experiments by Johnson~\textit{et al.}\cite{Johnson2008}. The type of packing algorithm and the selected size range of the granular gouge particles result in a quasi-uniform Particle Size Distribution (PSD). Figures~\ref{fig:fig2} shows the PSDs of both the roughness and granular gouge layers. 

The length of the system in the $X$ direction is 70$L_0$, while its thickness in the $Z$ direction is 5.46$L_0$. Periodic boundary conditions are employed in the $X$ direction. The two lateral sides of the medium in $Z$ direction are bounded by frictionless deformable walls with the same stiffness of the granular layer particles. 

The interaction of both the HCP and roughness layer particles are modeled by radial springs ~\cite{Place1999,Wang2006}. The inter-particle radial force is $F_r = K_r \cdot \Delta r$. $\Delta r$ is the difference between the inter-particle distance and the sum of the particle radii and $K_r=2.9775 \cdot 10^7$ $M_{0} \cdot {t_0}^{-2}$ is the radial compressional/tensional spring stiffness. The value of $K_r$ is chosen based on a calibration procedure developed by Schoepfer~\textit{et al.}~\cite{Schoepfer2009} to achieve a bulk Young modulus in typical ranges of rock materials for the particle assembly~\cite{Johnson2008,Mair2008}.
The granular gouge layer particles interact with each other and with the driving block/substrate particles via a repulsive Hookean spring with radial and tangential components that 
represents normal (to the contact plane) and frictional forces respectively~\cite{Place1999,Wang2006}. The radial component has a spring stiffness $K_r = 5.954.10^7$ $M_{0} \cdot {t_0}^{-2}$. The spring stiffness of the tangential component is $K_s = 5.954.10^7$ $M_{0} \cdot {t_0}^{-2}$ for all granular gouge particles. The frictional interaction among the granular gouge particles is implemented similar to the model proposed by Cundall $\&$ Strack~\cite{Cundall1979}. The tangential contact force is chosen as the minimum value of $K_s \cdot \Delta s$ and the Coulomb threshold value $\mu \cdot F_r$, at each time step. $\mu$ is the friction coefficient between the two particles’ surfaces and can be either static, $\mu_s$, or dynamic, $\mu_d$. We chose friction coefficient values of $\mu_{s} = \mu_{d} = 0.4$ to produce a macroscopic frictional behavior corresponding to quartz sand aggregates. The frictional interaction between the granular gouge particles and the roughness layers is modeled in the same way with the friction coefficients of $\mu_{static}=\mu_{dynamic}=0.7$. These values were adjusted based on a parametric study to enhance the stick-slip behavior by increasing the frictional interaction in the interface of the two layers. 

Each simulation run consists of two stages. During the first stage, the consolidation stage, no shear load is imposed and the granular layer is compressed by the vertical displacement of both the driving block and the substrate. The displacement continues until the applied normal stress on the granular layer equals the desired value of the confining pressure, $\sigma_n$. 

The second stage of each simulation run starts after the consolidation stage and consists in keeping the normal load constant on the driving block while applying a constant velocity of $V_{X,0} = 0.004 \frac{L_0}{t_0}$ to the top particles of the HCP layer of the driving block. The imposed velocity introduces a shear load to the granular system. A ramp protocol is employed for gradually increasing the shear velocity from $0$ to $V_{X,0}$ ~\cite{Griffa2011,Griffa2012,Griffa2013}. We identified different regions in the $\sigma_n-V_{X,0}$ parameter space where the system follows either a stick-slip dynamics or is in steady sliding mode. These two regimes are typical of dynamical regime for the granular layer~\cite{Aharonov1999}. 

In the specific case of the perturbed runs, \textit{i.e.}, when external vibration is applied, an additional boundary condition consists in imposing a displacement in the $Y$ direction for the bottom particles of the substrate.   
The temporal displacement of this boundary displacement is modelled as

\begin{eqnarray}
\label{YDisplACVibration}
	u_{Y}(t) = & A\cdot \Delta t \cdot \left[ \frac{\partial f}{\partial t}\left( t,t',T_{\nu},\tau \right) \cdot cos\left( \omega(t-t') - \frac{\pi}{2} \right) - \right. \nonumber \\
								& \left. - \omega \cdot f\left( t,t',T_{\nu},\tau \right) \cdot sin\left( \omega(t-t') - \frac{\pi}{2} \right) \right]
\end{eqnarray}
where

\begin{equation}
\label{AmplModACVibration}
	f\left( t,t',T_{\nu},\tau \right) \equiv \frac{1}{2} \cdot \left[ tanh\left( \frac{t - t'}{\tau} \right)  - tanh\left( \frac{t - (t' + T_{\nu})}{\tau} \right) \right].
\end{equation}

\begin{figure}[h]
\includegraphics [width=0.45\textwidth]{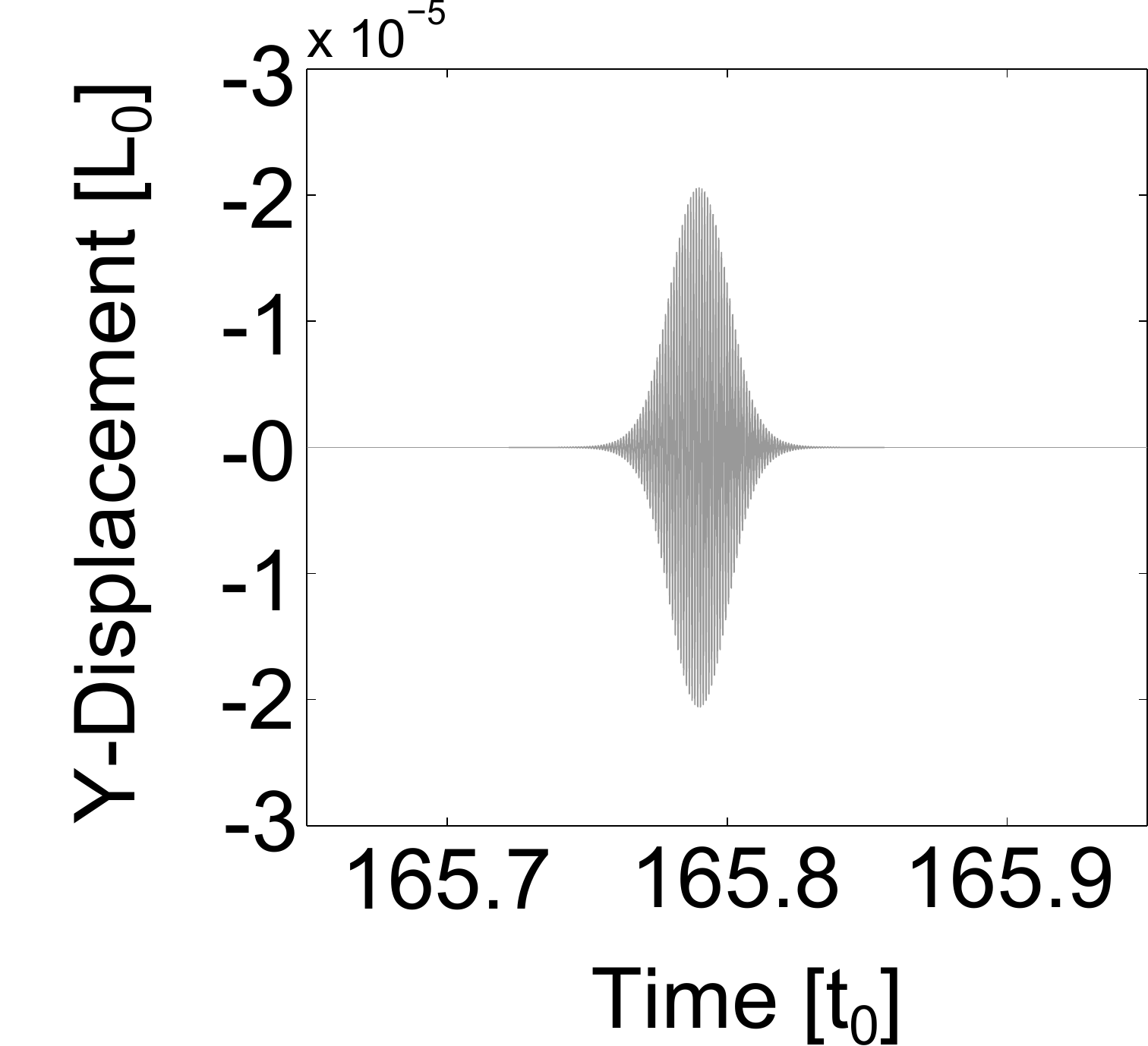}
\caption{\label {fig:modelamplitude} Example of the displacement imposed to the bottom layer of the substrate in the $Y$ direction, $u_y$, as an AC vibration source at the boundary of the system. It represents a harmonic oscillation at frequency $f_0 =\ 1\ kHz$ with amplitude
modulation given by a Gaussian-like signal. The peak to peak amplitude of this AC displacement $u_{y,PP} = 4.0.10^{-5}L_0$ is quite small compared to the largest particle size within the medium }
\end{figure}

In Eqs. \ref{YDisplACVibration} and \ref{AmplModACVibration}, $t = m \cdot \Delta t$, $\forall\ m=0,1,... $, is discretized time and $\Delta t$ is the simulation time step.
Eq. \ref{YDisplACVibration} represents a sinusoid with angular frequency $\omega = 2\pi \cdot f_{0}$, with $f_0 =\ 1\ kHz$, whose amplitude is modulated in time by a waveform with a Gaussian-like shape, given by Eq. \ref{AmplModACVibration}.
In Eq. \ref{YDisplACVibration}, $t'$ represents a phase shift term for centering the temporal window of the vibration at different times during the stick-slip dynamics.
$\tau = 0.01$ and $T_{\nu} = 0.02$, in Eq. \ref{AmplModACVibration}, play respectively the role of a rising/decaying time constant and width for the displacement waveform.
In Eq. \ref{YDisplACVibration}, $A$ is the the vibration peak amplitude value. Figure~\ref{fig:modelamplitude} shows an example of the displacement waveform applied within the time interval $[165.75; 165.85]$.

For the implementation of the model, we used the open source code ESyS-Particle, developed at and maintained by the Earth Systems Science Computational Center of the University of Queensland,
Brisbane, Australia. ESyS-Particle solves  Newton's equations of motion for the center of mass of each particle are solved by a first order, explicit finite difference scheme and for the rotation angles about the center of mass by a finite difference rotational leapfrog algorithm\cite{Wang2009c}.
The finite difference time step $\Delta t =\ 15 \cdot 10^{-6}$ is small enough to guarantee numerical stability and to satisfy the sampling theorem for a vibration signal with maximum frequency $f_{max} =\ 2 \cdot 10^{5}$,
which is approximately the maximum sound frequency of vibration in the laboratory experiments by Johnson \textit{et al.} \cite{Johnson2008}. 

\section{Results}
We define the macroscopic friction coefficient, $\mu$, as the shear stress of the granular layer divided by its confining stress and we monitor its time-variation to investigate the behavior of the granular layer. The total kinetic energy of the granular layer is another variable that we use to investigate the state of the granular layer. The total kinetic energy of each j-th particle belonging to the granular layer, $K_j$, is defined as $K_j= K_j^{trans} + K_j^{rot}$, where $K_j^{trans}$ is the j-th particle translational kinetic energy and $K_j^{rot}$ is its rotational kinetic energy. We define the total kinetic energy for the overall granular layer as ${K_{tot}\equiv \sum_{j = 1,...,M} K_j}$, with $M$ the total number of granular layer particles. 

We performed simulations at two confining pressures of $\sigma_{n} = 4$ MPa and $40$ MPa. The confining pressure of $\sigma_{n} = 4$ MPa results in steady sliding of the sheared granular layer, while the confining pressure of $\sigma_{n} = 40$ MPa shows stick-slip dynamics. Both of the confining pressures are in the range observed on geological fault settings as well as in experimental setups (a few to a few hundreds MPas). The simulations that are not exposed to any vibration are called ``reference'' run, while the simulations with vibration are called ``perturbed'' run. For each confining pressure, several separate ($\textit{i.e.}$ once for each perturbed run) vibrations at different shear stress levels are applied. Different vibration amplitudes ranging from $\{1, 5, 10, 20, 30, 40, 50, 60, 70, 80, 90, 100,\\ 200, 300, 400\}.10^{-7}L_0$ are used at each vibration interval for the two confining pressures. The response of the granular layer in the form of a frictional weakening during boundary vibration is studied in detail in this article. 

\begin{figure}
\includegraphics[width=0.9\textwidth]{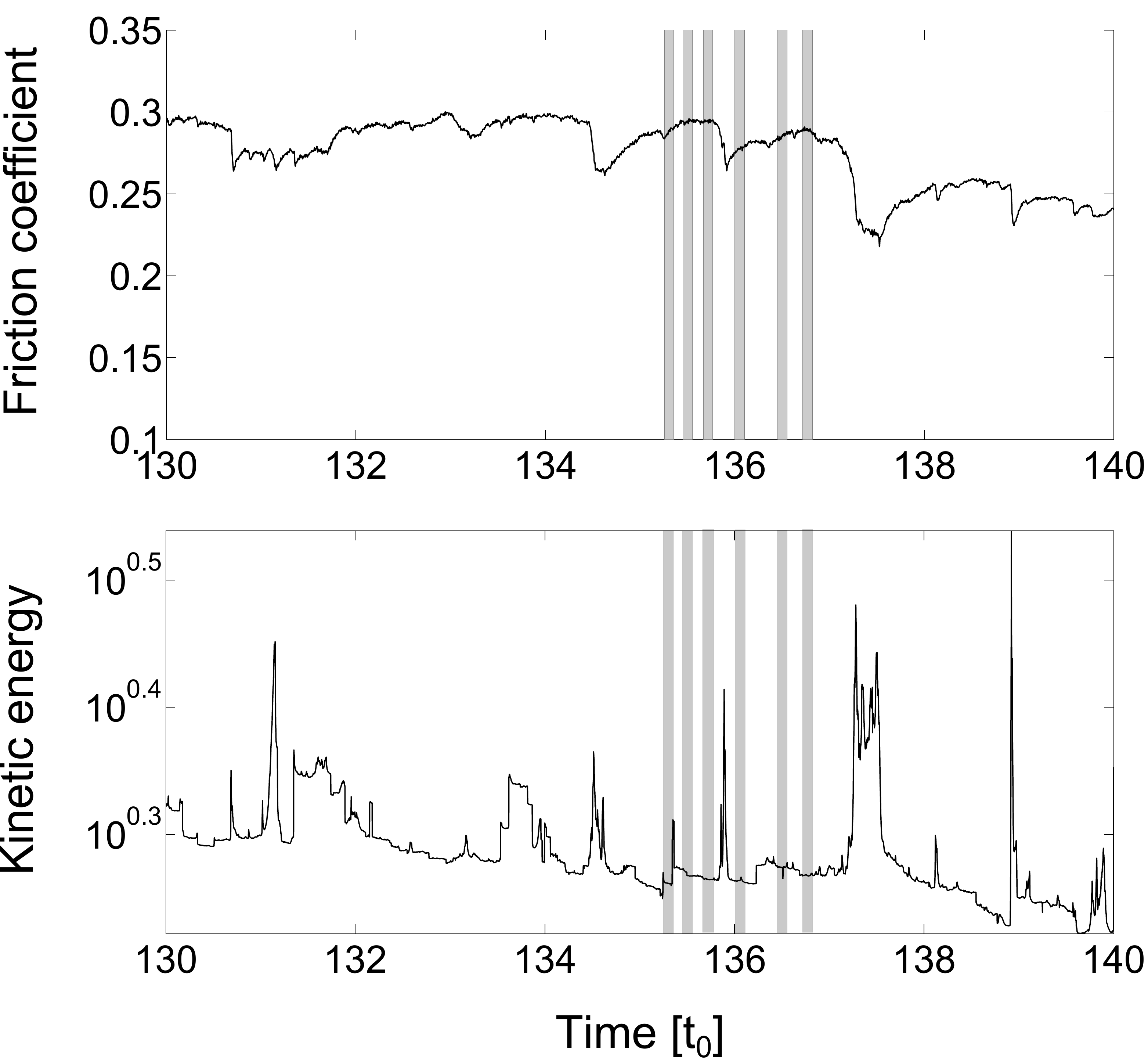}
\caption{Friction coefficient (top panel) and kinetic energy (bottom panel) signals for the reference run at $\sigma_n = 4$ MPa. Vibration intervals are illustrated with vertical dashed lines.}
\label{fig:fig3}
\end{figure}

\begin{figure}
\includegraphics[width=0.9\textwidth]{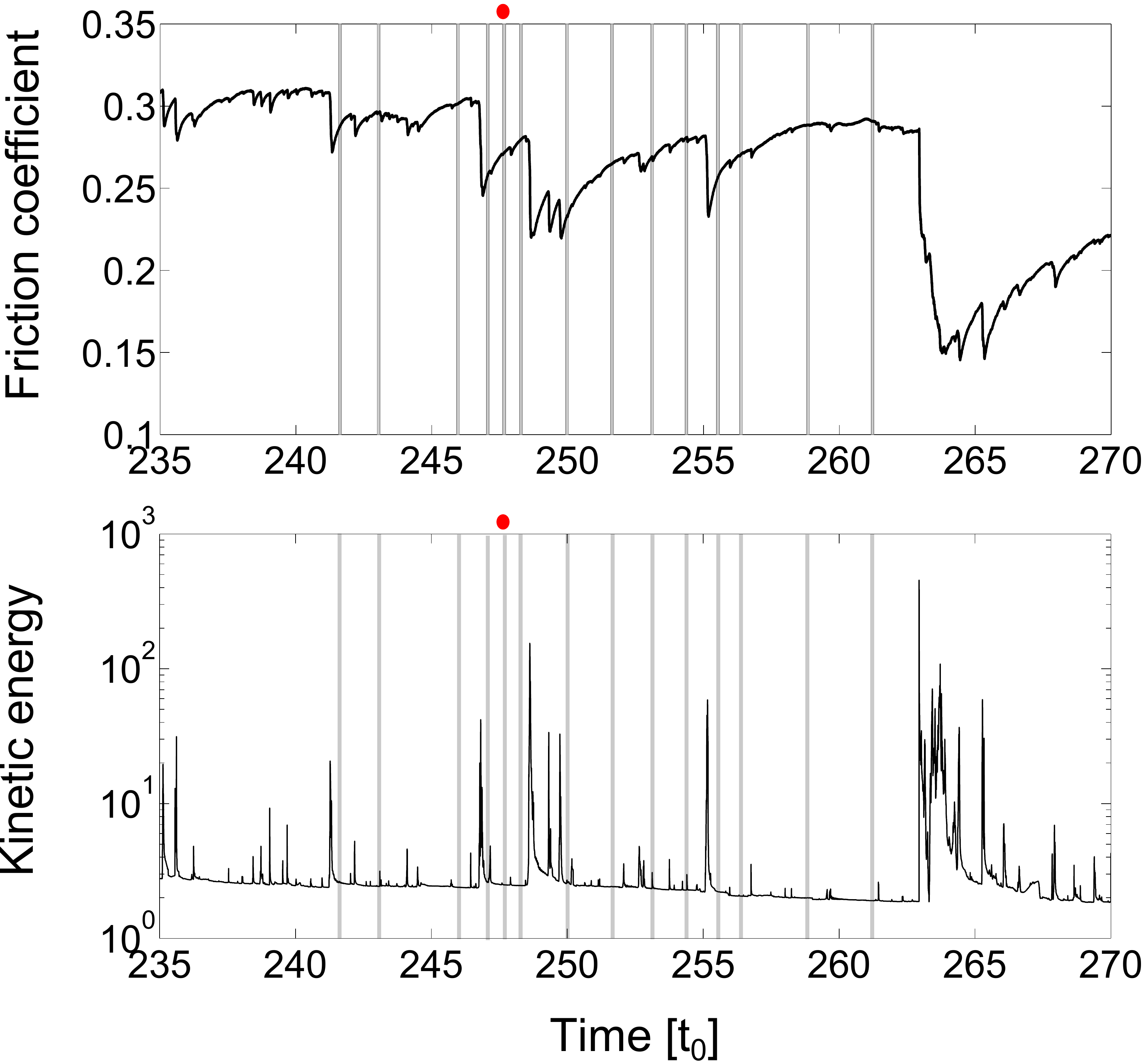}
\caption{Friction coefficient (top panel) and kinetic energy (bottom panel) signals for the reference run at $\sigma_n = 40$ MPa. Vibration intervals are illustrated with vertical dashed lines.}
\label{fig:fig4}
\end{figure}

Figure~\ref{fig:fig3} shows the friction coefficient and kinetic energy signals for the granular layer confined at $\sigma_n = 4$ MPa. The friction coefficient signal varies between 0.2 to 0.3 and shows some fluctuations corresponding to small instabilities in the layer. The energy release of these friction coefficient fluctuations is a fraction of the background (ambient) energy. We take this behavior to characterize the steady-sliding regime. The vibrations (6 intervals) are applied at different shear stress levels and are shown as vertical dashed lines in this figure. The Friction coefficient and kinetic energy signals for the granular layer at $\sigma_n = 40$ MPa are plotted in figure~\ref{fig:fig4}. The friction coefficient signal varies between 0.1 to 0.3 and shows clear evidence of a stick-slip regime characterized by series of long-lasting increases followed by sudden, fast drops. The kinetic energy jumps during slip events are 2-3 orders of magnitude larger than the background (ambient) energy level. Therefore, the behavior at confining pressure of $\sigma_n = 40$ MPa corresponds to regular stick-slip dynamics of the granular layer. The vibrations (14 intervals) are applied at different shear stress levels and are shown as vertical dashed lines in the figure. 

\begin{figure*}
\includegraphics[width=1.105\textwidth]{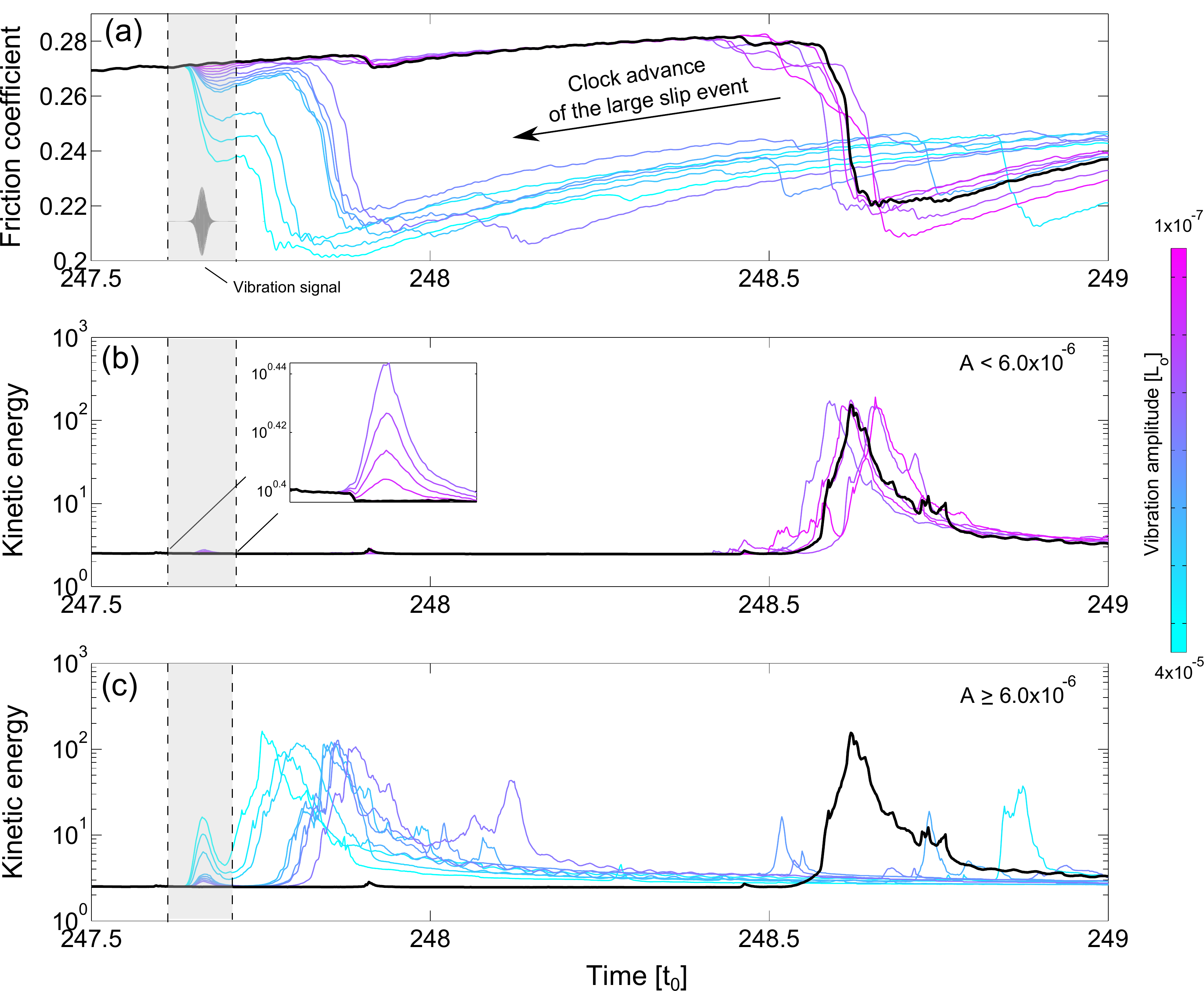}
\caption{Macroscopic signatures of the effect of increasing vibration amplitudes: time series of (a) friction coefficient, (b) total kinetic energy (for small vibration amplitudes, $A < 6.10^{-6}L_0$), and (c) total kinetic energy (for larger vibration amplitudes, $A \geq 6.10^{-6}L_0$). The vibration interval is indicated by vertical dashed lines and shadowed area in all panels. This interval is also shown with a red marker in figure~\ref{fig:fig4}.}
\label{fig:fig5}     
\end{figure*}

Figure~\ref{fig:fig5}-a shows an example of the friction coefficient time-series, $\mu$, for the reference (black line) and the perturbed runs with a range of vibration amplitudes for $\sigma_n = 40$ MPa. The reference run time-window shows the stick-phase of a large slip event happening at $t = 248.7 t_0$. The vibration interval is illustrated with vertical dashed lines in the figure. This interval is also shown with a red marker in figure~\ref{fig:fig4}. Large enough vibration amplitudes induce a reduction of the friction coefficient during the vibration interval. The reduction of friction coefficient becomes more significant at larger vibration amplitudes. In addition, vibration amplitudes of $A \geq 6.10^{-6}L_0$ induce a noticeable clock-advance of the upcoming large slip event in the perturbed runs compared to the reference run, while amplitudes of $A < 6.10^{-6}L_0$ do not change the time of large slip event. An in depth study of this clock advance effect is not the scope of this article and will be investigated elsewhere. The focus of this article is studying the weakening of the frictional behavior during vibration interval that we termed it ``frictional weakening event'' in the rest of the article.

The kinetic energy signal for the reference and perturbed runs are shown in figs.~\ref{fig:fig5}-b (for vibration amplitudes, $A,<6.10^{-6}L_0$) and ~\ref{fig:fig5}-c (for $A \geq6.10^{-6}L_0$). The vibration interval is illustrated with vertical dashed lines and a shadowed area. The kinetic energy signal attains a background value during the stick-phase that corresponds to the constant shearing of the dense granular layer. During a large slip events, kinetic energy abruptly increases compared to the background level. This corresponds to a transfer of energy from the elastic potential energy to the kinetic one. In the perturbed runs and during the vibration interval, kinetic energy slightly and slowly increases which corresponds to the frictional weakening event induced by the boundary vibration. A zoom (of the vibration interval) is given in the inset of the kinetic energy signal in fig.~\ref{fig:fig5}-b. Vibration amplitudes of $A \leq 1.10^{-6}L_0$ produce only small fluctuations in the kinetic energy signal, therefore we take $A = 1.10^{-6}L_0$ to be the threshold for inducing frictional weakening. The other larger vibration amplitudes induce frictional weakening events that are completely visible in the kinetic energy signal. The increase in the kinetic energy corresponding to these frictional weakening events develops more slowly and is noticeably smaller compared to regular slip events. 

\begin{figure}
\includegraphics[width=1.105\textwidth]{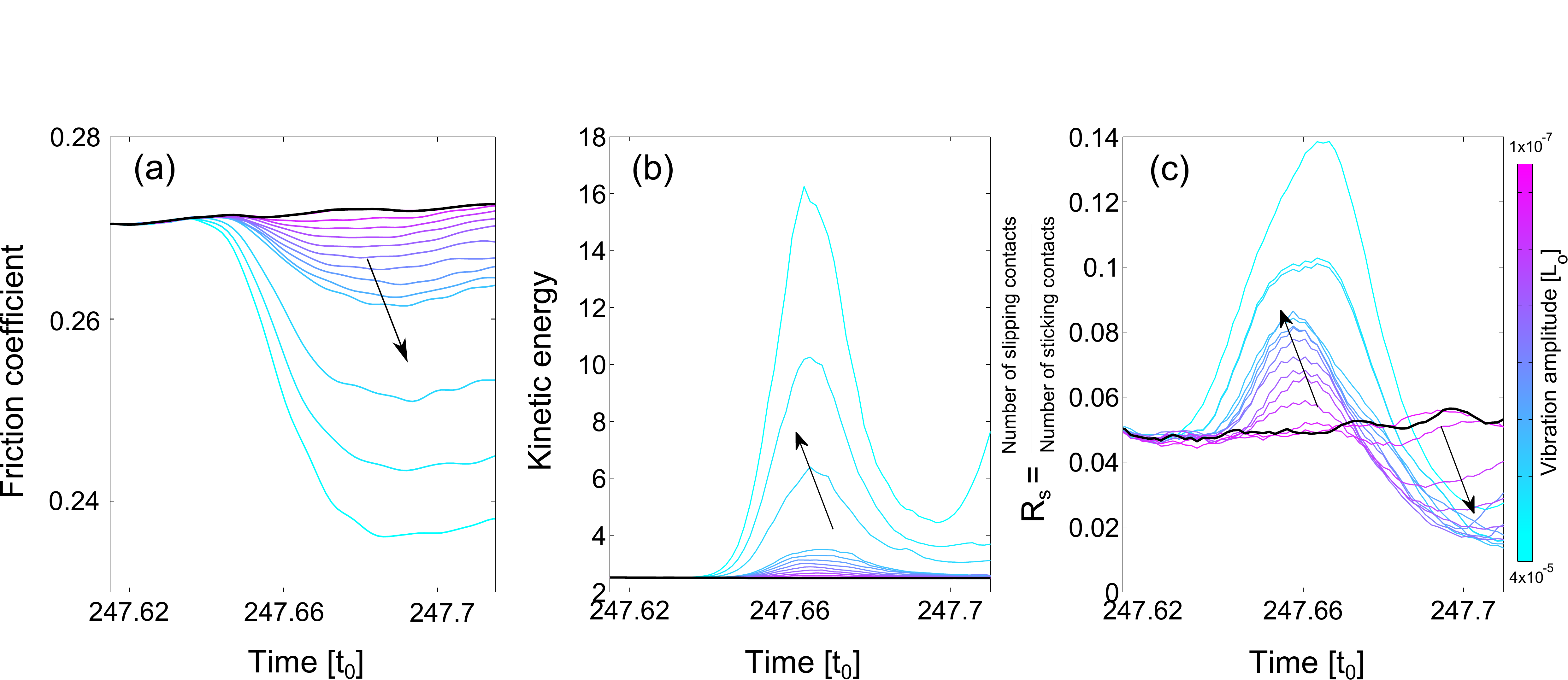}
\caption{(a) Friction coefficient time series, (b) Kinetic energy time series, (c) Ratio of slipping to sticking contacts for the reference and perturbed runs. The arrow indicates the increase of the vibration amplitude. The colormap is the same that is used in fig.~\ref{fig:fig5} and the black line corresponds to the reference run. The time-span in the figure corresponds to the vibration interval.}
\label{fig:fig5_1}     
\end{figure}

To track the boundary vibration effect at the grain-scale, we show in figure~\ref{fig:fig5_1}-c the ratio of the number of slipping to sticking contacts, $R_s$, in the granular gouge layer for the reference and perturbed runs for different vibration amplitudes. Panels (a) and (b) of figure~\ref{fig:fig5_1} shoe friction coefficient and kinetic energy time series during the vibration interval. Slipping contacts are those contacts that reach the grain-scale dynamic friction, $\mu_{dynamic}$, while sticking contacts are those that are still in the grain-scale static friction, $\mu_{static}$. The time span of figure~\ref{fig:fig5_1} corresponds to the vibration interval. Evolution of the $R_s$ starts immediately upon the vibration application, therefore it starts even with quite small boundary perturbations. Additionally, the increase of the $R_s$ starts earlier (within the rising time of the vibration signal) than the kinetic energy signal. The time-lag between these two (the increase of $R_s$ and evidence of frictional weakening in the kinetic energy signal) is due to the fact that an increase of $R_s$ larger than a certain limit is essential for the mobilization of grains and releasing measurable energy. The $R_s$ reaches a maximum at about the peak of the vibration displacement signal. A profound increase in the $R_s$ is observed for large vibration amplitudes, $A \geq 6.10^{-6}L_0$, while there is no significant change of the $R_s$ for $A \leq 1.10^{-6}L_0$. This indicates that vibration amplitudes larger than a threshold cause an irreversible change in the contact networks in the form of significant amount of grain contact rearrangement. The result of this irreversible evolution is the frictional weakening of the layer. 
The decreasing trend continues until the end of the vibration interval. Except for $A \leq 1.10^{-6}L_0$, the $R_s$, at the end of vibration interval, goes below its initial value before that vibration applied. This is due to the decrease of the friction coefficient level of the granular layer which itself was induced by the evolution of grain contact network.  The boundary vibration influence appears to be longer-lived in the contacts network than in the kinetic energy signal. 


In the next two sections, we present results on the influence of the vibration amplitude and the shear stress level the vibration is applied on
the size of the frictional weakening events. The drop in the friction coefficient is used as a measure of the event size. We use the total released kinetic energy as a measure of the event size. The total released kinetic energy during an event, $E$,  is defined as $E=\sum _{i=1} ^{N} (K_{tot}-K_{tot,0}) \cdot\dot{\gamma} \cdot \Delta t$. $K_{tot,0}$ is the background value of $K_{tot}$ in the reference run. $\dot{\gamma}$ is the shear strain rate of the driving block, calculated as the temporal derivative of the ratio between the driving block top layer's displacement and the granular layer thickness. $\Delta t$ is the simulation time step. The sum is performed over the total number of continuous values of $K_{tot}$ during the frictional weakening event (within vibration interval).  

\subsection{Effect of vibration amplitude}

\begin{figure}
\includegraphics[width=0.85\textwidth]{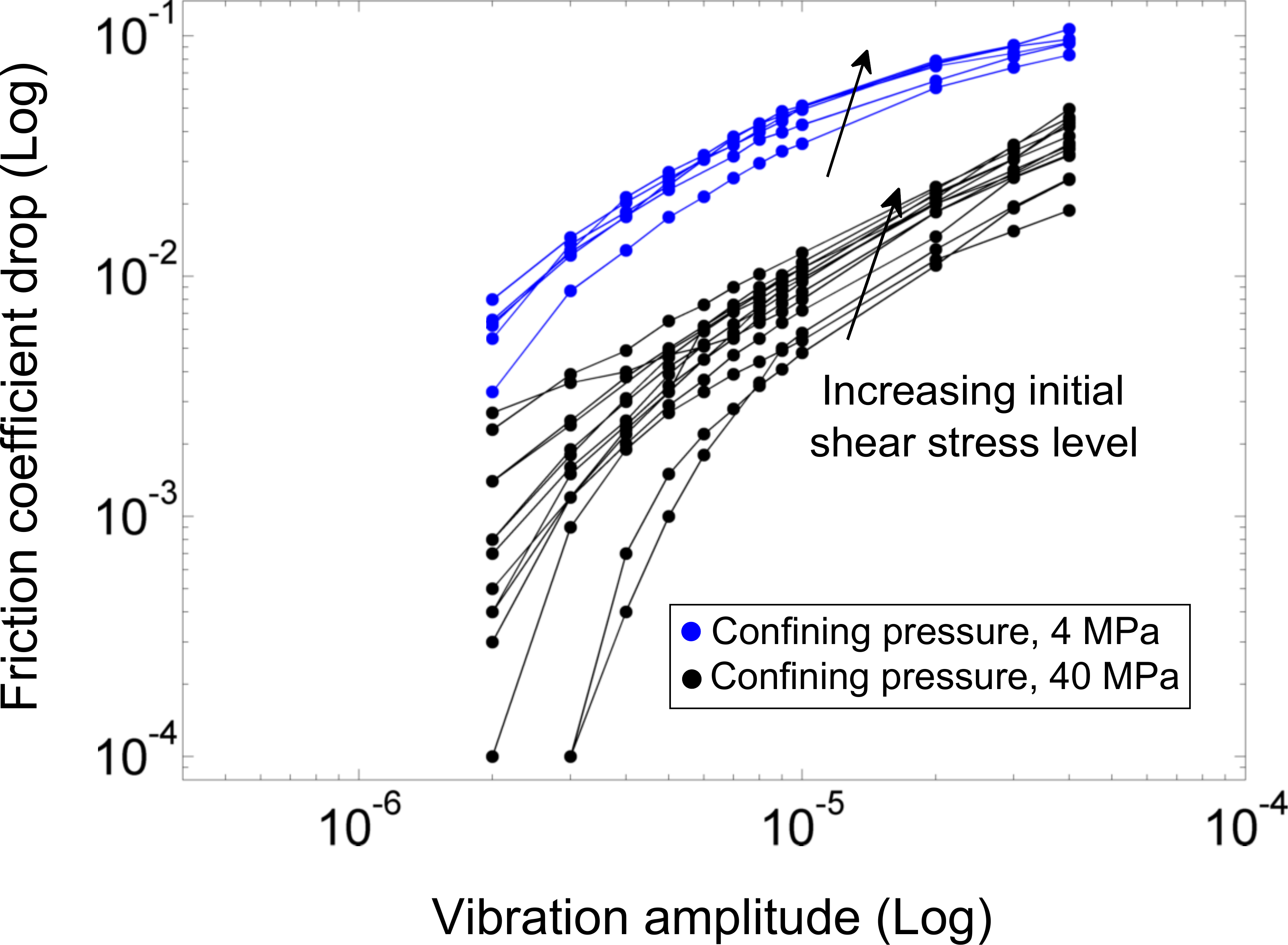}
\caption{Friction coefficient drop as a measure of the frictional weakening event size for two different confining pressures of $\sigma_n = 4$ and $40$ MPa and different vibration amplitudes. Different lines for each confining pressures refer to the different shear stress levels where the vibration is applied.}
\label{fig:fig6}  
\end{figure}

Figure~\ref{fig:fig6} shows the drop in friction coefficient associated to frictional weakening events for different vibration amplitudes. A clear observation is that the friction drop size increases by increasing the boundary vibration amplitude. In addition, we could not see any measurable (beyond fluctuation) change in the friction coefficient signal for vibration amplitudes of $A \leq 1.10^{-6}L_0$. The friction drop increases when vibration is applied at a higher shear stress level.

%

\begin{figure}
\includegraphics[width=0.85\textwidth]{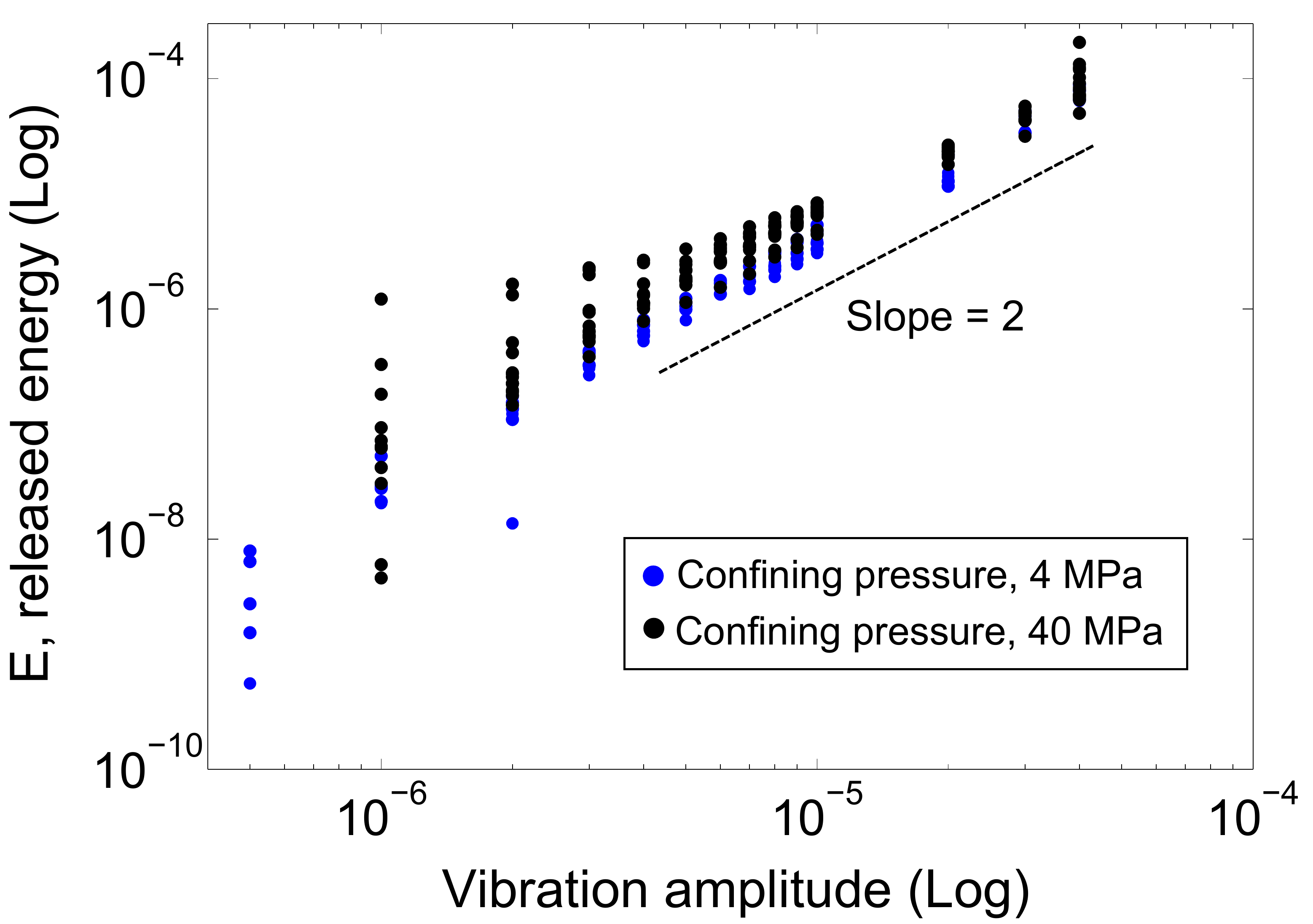}
\caption{Total kinetic energy release as a measure of the frictional weakening event size for two different confining pressures of $\sigma_n = 4$ and $40$ MPa and different vibration amplitudes. Different lines for each confining pressures refer to the different initial shear stress levels where the vibration is applied.}
\label{fig:fig8}       
\end{figure}

The kinetic energy release of the frictional weakening events for different vibration amplitudes is presented in figure~\ref{fig:fig8}. Larger vibration amplitude results in larger energy release. 
This fact is not the simple consequence of a direct transfer of energy from the vibration source to the friction drop events. Indeed, we calculated the amount of work done to the granular layer by the boundary vibration and determined that it is in the order of $10^{-5}$ to $10^{-2}$ of the released kinetic energy. This indicates that the increase in the released kinetic energy does not simply mirror the increase in the energy input into the system by the applied vibration but it mirrors an increase of the vibration efficiency in unlocking the particle contacts and facilitating particle rearrangements and mobilizations. 
There are fewer fluctuations in the amount of kinetic energy release at different shear stress levels as the vibration amplitude increases.  Furthermore, vibration amplitudes of $A \leq 1.10^{-6}L_0$ and $A \leq 5.10^{-7}L_0$ do not induce a measurable energy release for the confining pressures of $\sigma_n = 40$ and $4$ MPa, respectively.

\subsection{Effect of shear stress level}

\begin{figure}
\includegraphics[width=0.85\textwidth]{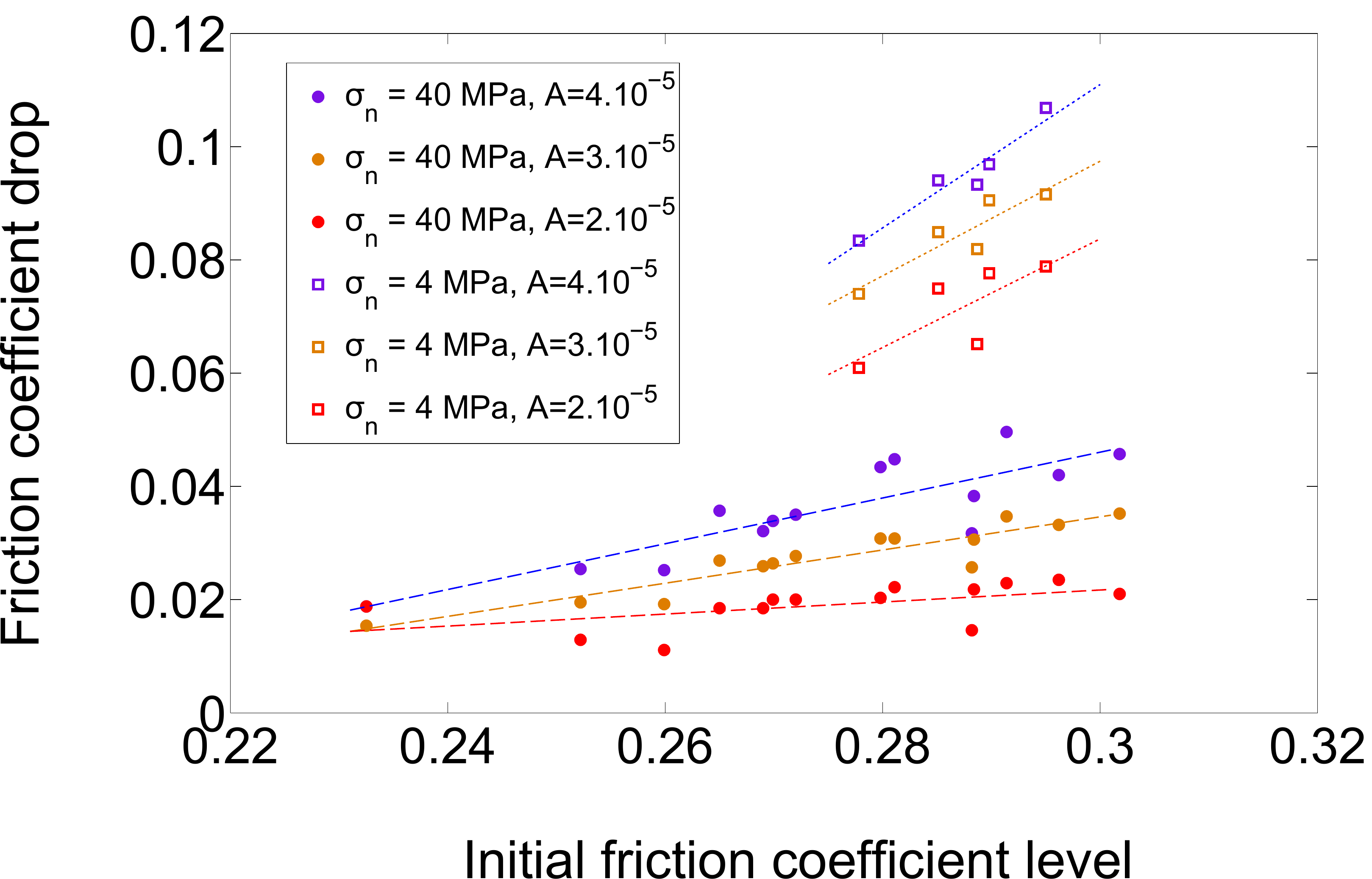}
\caption{Friction coefficient drop as a measure of the frictional weakening event size for the simulations at $\sigma_n = 4$ and $40$ MPa. The results correspond to the three largest vibration amplitudes.}
\label{fig:fig7}    
\end{figure}

We investigate in this section the influence of the shear stress level at which the vibration is applied. Figure~\ref{fig:fig7} shows the friction coefficient drop associated with the friction weakening events versus the shear stress level for the three largest vibration amplitudes. The figure shows that for both of the confining pressures, friction drop size of the friction weakening event is on average larger at higher shear stress level. Influence of the shear stress level is more significant for simulation at $\sigma_n = 4$ MPa, which is due to the fact that the medium at this pressure is more mobile and easier to perturb. Furthermore, large vibration amplitudes increase the influence of the shear stress level on the size of frictional weakening event for simulations at $\sigma_n = 40$ MPa, while they have no significant effect in this sense on the size of frictional weakening events for simulations at $\sigma_n = 4$ MPa. This difference could be due to the way the vibration displacement and the confining stress control are implemented in our simulations. After the consolidation stage is finished, the confining stress is only controlled via the upper boundary, to avoid its interference with the vibration displacement applied via the lower boundary. 

\begin{figure}
\includegraphics[width=0.85\textwidth]{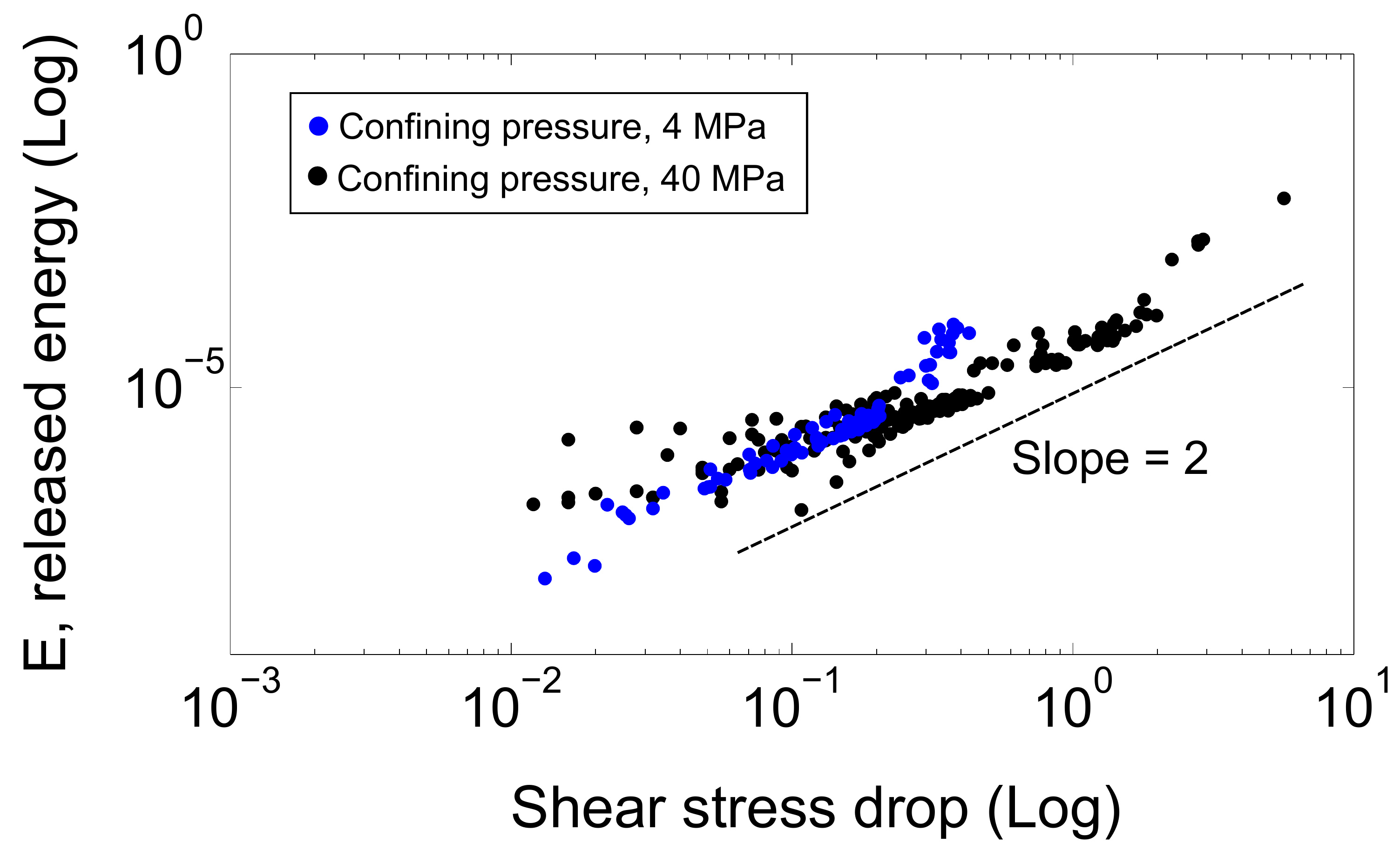}
\caption{Variation of the total kinetic energy release with the associated shear stress drop drop for frictional weakening events for two confining pressures of $\sigma_n = 4$ and $40$ MPa.}
\label{fig:fig10}       
\end{figure}

As a last point, figure~\ref{fig:fig10} shows the kinetic energy release versus the shear stress drop of all friction weakening events. It appears that the kinetic energy release is proportional to the shear stress drop squared, irrespective of the confining stress of the granular layer. The scaling of kinetic energy release with the stress drop is in agreement with the slip-weakening friction law prediction~\cite{Ruina1983,Palmer1973,Ida1972}. 


\section{Discussions}
We showed that the ratio of slipping to sticking contacts, $R_s$, increases significantly in correspondence of large vibration amplitudes and allows for noticeable grain contact network rearrangements. These induced rearrangements enhance particles mobilization and cause frictional weakening and kinetic energy release. This observation is consistent with the proposed hypothesis by Jia~\textit{et al.} about the evolution of the contact network in the presence of appreciable acoustic perturbation where the granular medium arrives at an irreversible regime and elastic weakening occurs as a result of the sound-matter interaction~\cite{Jia2011}. Furthermore, we showed that an amplitude threshold exists for triggering of frictional weakening events and this threshold is larger for higher confining stresses. This is also in agreement with the recent experimental observation by Jia~\textit{et al.} where they found that the acoustic fluidization threshold increases by increasing the confining pressure of the granular medium~\cite{Jia2011}. 

We showed in fig.~\ref{fig:fig5}-d that the frictional weakening events are much smaller and slower than regular large slip events in our numerical simulation. This makes them to be similar to the experimental ``slow slip" events observed in the laboratory by Johnson \textit{et al.}\cite{Johnson2012}. The amplitude dependence of frictional weakening events size is further in accordance with their experimental observations~\cite{Johnson2012}. 

\section{Conclusion}
We have studied vibration-induced frictional weakening phenomenon in a dense sheared granular layer by 3D Discrete Element Method (DEM) modeling. The frictional weakening was evaluated based on its associated friction coefficient drop as well as its kinetic energy release. We found that friction coefficient drop and kinetic energy release scales with vibration amplitude, \textit{i.e.} larger vibration amplitude results in larger frictional weakening events. The ratio of slipping to sticking contacts is used to explain the grain-scale mechanism of the frictional weakening phenomenon. This ratio increases significantly in correspondence of large vibration amplitudes and allows for noticeable grains contact network rearrangements, particles mobilization and consequently kinetic energy release. In addition to characterizing the physics of vibration induced weakening, a primary goal of this study is to advance the understanding of the physics of the Dynamic Earthquake Triggering (DET) phenomenon. The existence of a threshold for the vibration amplitude below which no appreciable grains contact network rearrangement and immediate frictional weakening occurs and therefore has no significant influence on the upcoming large slip event is in agreement with the laboratory (Johnson {\textit et al.} \cite{Johnson2012}) and field-scale (Gomberg $\&$ Johnson \cite{Gomberg2005}) observations for DET.

\begin{acknowledgements}
We thank D. Weatherley and S. Abe for support during the implementation of our model in the
ESyS-Particle code and D. Passerone and C. Pignedoli for the help related with
the use of the High Performance Computing cluster, Ipazia, at Empa.
Our work has been supported by the Swiss National Science Foundation (projects No. 206021-128754 and No. 200021-135492)
and by the LDRD Program (Institutional Support) at the Los Alamos National Laboratory, Dept. of Energy, USA.
\end{acknowledgements}

\bibliographystyle{nature}  
\bibliography{libraryPRE}

\end{document}